\documentclass[acmtog]{acmart}

\usepackage{booktabs} 

\usepackage{arydshln}

\usepackage{xcolor}
\usepackage{tikz}
\usepackage{multirow}
\usepackage{enumitem}

\usepackage[ruled]{algorithm2e} 

\SetAlFnt{\small}
\SetAlCapFnt{\small}
\SetAlCapNameFnt{\small}
\SetAlCapHSkip{0pt}

\begin{document}
\title{CrowdVLA: Embodied Vision-Language-Action Agents for Context-Aware Crowd Simulation}

\author{Juyeong Hwang}
\authornote{These authors contributed equally to this work.}
\affiliation{%
  \institution{IIIXR Lab, Korea University}
  \city{Seoul}
  \country{South Korea}
}
\email{05judy02@korea.ac.kr}

\author{Seong-Eun Hong}
\authornotemark[1]
\affiliation{%
  \institution{IIIXR Lab, Korea University}
  \city{Seoul}
  \country{South Korea}
}
\email{seong\_eun@korea.ac.kr}

\author{Jinhyun Kim}
\affiliation{%
  \institution{IIIXR Lab, Korea University}
  \city{Seoul}
  \country{South Korea}
}
\email{rlawlsgus0623@korea.ac.kr}

\author{JaeYoung Seon}
\affiliation{%
  \institution{IIIXR Lab, Kyung Hee University}
  \city{Seoul}
  \country{South Korea}
}
\email{cogongnam@khu.ac.kr}

\author{Giljoo Nam}
\affiliation{%
  \institution{Meta}
  \city{United States}
  \country{United States}
}
\email{giljoonam@khu.ac.kr}

\author{Hanyoung Jang}
\affiliation{%
  \institution{NC AI}
  \city{Seoul}
  \country{South Korea}
}
\email{ncdrjang@ncsoft.com}

\author{HyeongYeop Kang}
\authornote{Corresponding author.}
\affiliation{%
  \institution{IIIXR Lab, Korea University}
  \city{Seoul}
  \country{South Korea}
}
\email{siamiz\_hkang@korea.ac.kr}

\renewcommand{\shortauthors}{Hwang, Hong, Seon, and Kang}

\begin{teaserfigure}
  \includegraphics[width=\textwidth]{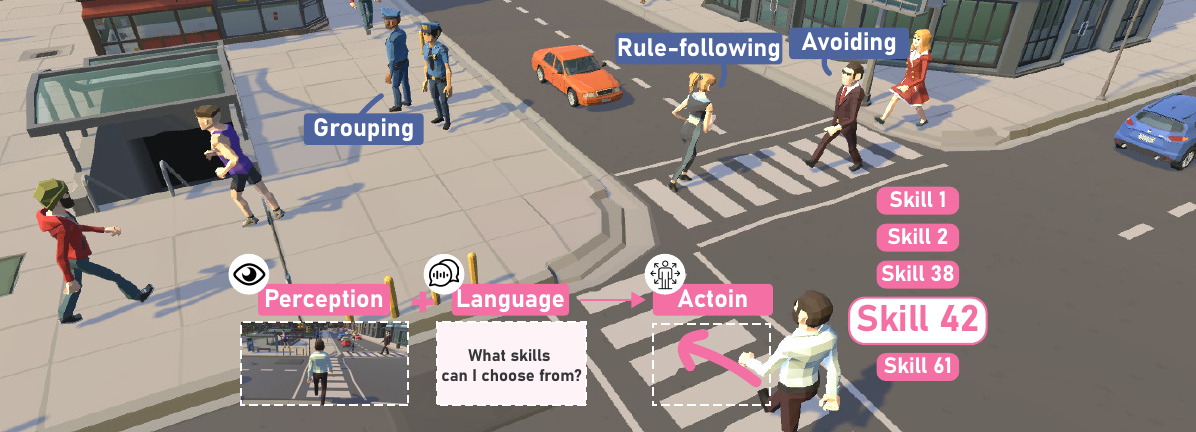}
  \caption{CrowdVLA conditions motion skill actions on visual input and language-based instructions, enabling agents to reason about scene semantics and social norms before selecting appropriate behaviors across diverse environments.}
\end{teaserfigure}

\begin{abstract}
Crowds do not merely move; they decide.
Human navigation is inherently contextual: people interpret the meaning of space, social norms, and potential consequences before acting. Sidewalks invite walking, crosswalks invite crossing, and deviations are weighed against urgency and safety. Yet most crowd simulation methods reduce navigation to geometry and collision avoidance, producing motion that is plausible but rarely intentional.
We introduce CrowdVLA, a new formulation of crowd simulation that models each pedestrian as a Vision—Language--Action (VLA) agent. Instead of replaying recorded trajectories, CrowdVLA enables agents to interpret scene semantics and social norms from visual observations and language instructions, and to select actions through consequence-aware reasoning.
CrowdVLA addresses three key challenges--limited agent-centric supervision in crowd datasets, unstable per-frame control, and success-biased datasets--through: (i) agent-centric visual supervision via semantically reconstructed environments and Low-Rank Adaptation (LoRA) fine-tuning of a pretrained vision--language model, (ii) a motion skill action space that bridges symbolic decision making and continuous locomotion, and (iii) exploration-based question answering that exposes agents to counterfactual actions and their outcomes through simulation rollouts. Our results shift crowd simulation from motion-centric synthesis toward perception-driven, consequence-aware decision making, enabling crowds that move not just realistically, but meaningfully.

\end{abstract}


\maketitle

\section{Introduction}
\label{sec:introduction}

Simulating crowds that autonomously navigate complex environments is a long-standing challenge in computer graphics. Realistic crowd behavior extends beyond collision-free motion toward a goal. 
Human navigation is inherently context dependent: individuals perceive their surroundings, interpret semantic and social cues, and act accordingly. 
Navigation decisions are shaped not only by physical constraints, but also by how spaces are socially interpreted: for example, preferring sidewalks over grass or crossing streets at crosswalks. As a result, realistic crowd behavior emerges from individual decisions conditioned on the interpreted use of space. 

Despite steady progress, most existing crowd simulation methods fall short of this perspective. Macroscopic models~\cite{guy2010pledestrians, helbing1995social, guy2011simulating, qiu2010modeling, lee2007group, guy2009clearpath, van2008reciprocal} reproduce aggregate flow patterns, while microscopic models~\cite{guy2010pledestrians, helbing1995social, guy2011simulating, qiu2010modeling, lee2007group, guy2009clearpath, van2008reciprocal} focus on local interactions such as collision avoidance and grouping. Both primarily frame navigation as motion governed by physical feasibility and nearby agents. Recent learning-based approaches~\cite{lee2018crowd, charalambous2023greil, panayiotou2022ccp, zheng2019improved, wei2018learning} inherit this formulation, replacing explicit rules with learned policies but still treating the environment as a geometric constraint space. Perception is typically used to detect obstacles and free space. This limitation is reinforced by commonly used datasets~\cite{lerner2007crowds, pellegrini2009you, yi2015understanding, robicquet2016learning}, which mainly provide 2D trajectories and bias models toward imitation of observed motion. Consequently, prior work produces locally plausible motion but struggles to model individuals whose behavior depends on how space is intended to be used.

We argue that crowd simulation should instead be formulated as a Vision--Language--Action (VLA) problem at the level of individual agents. Crowd navigation naturally requires agents to perceive complex scenes, interpret semantic and social cues, and select actions whose consequences unfold over time, closely aligning with recent VLA formulations in embodied AI. However, directly adopting VLA paradigms for crowds presents key challenges: agent-centric image--language--action supervision is scarce, per-frame action prediction is unstable for continuous locomotion, and trajectory datasets overwhelmingly capture only successful behavior, offering little supervision for reasoning about failures or unsafe alternatives.

To address these issues, we introduce three key design choices. First, we construct semantically structured environments with in Unity and re-render agent-centric third-person observations, enabling scalable Low-Rank Adaptation (LoRA)-based fine-tuning~\cite{hu2022lora} of a pretrained Vision–Language Model (VLM) for crowd navigation. 
Second, we discretize motion into short-horizon trajectory-based motion skills spanning 20 frames, derived from real crowd data. Selecting among motion skills, rather than regressing per-frame controls, provides stable and temporally coherent action grounding while preserving realistic movement.
Third, to overcome success bias in crowd datasets, we introduce exploration-based Question Answering (QA) supervision that explicitly rolls out counterfactual actions in the environment to observe their outcomes, such as collisions, norm violations, or inefficient detours. This trains agents to reason not only about which action to take, but why certain actions lead to undesirable consequences.

Together, these components enable context-aware, outcome-aware decision making for crowd simulation. Our contributions are summarized as follows:
\begin{itemize}
    \item \textbf{Introducing VLA agents for crowd simulation.} Crowd behavior is modeled with an end-to-end VLA policy that embeds semantic understanding and social reasoning directly into individual agent decisions.
    \item \textbf{Agent-centric supervision for VLA training.} Semantically structured Unity environments provide scalable agent-centric observations for fine-tuning pretrained VLMs.
    \item \textbf{Stable action grounding via motion skill abstraction.} Short-horizon motion skills derived from real trajectories enable stable and coherent action selection.
    \item \textbf{Exploration-based QA for learning beyond success.} Exploration-based supervision enables agents to reason about rare, unsafe, or unseen outcomes absent from success-biased datasets. 
\end{itemize}

\section{Related Work}
\label{sec:related_work}

\subsection{Crowd Simulation in Complex Environments}
The interaction between agents and their surroundings is central to plausible crowd simulation. While early methods successfully incorporated the environment as physical geometry, capturing the semantic affordances and context-dependent social conventions that drive human navigation remains a significant challenge.

\paragraph{Geometric and Planning-Based Approaches.}
Traditional crowd simulation frameworks typically rely on explicit environmental representations to handle complex scenes. Continuum methods, for instance, utilize semantically partitioned regions or globally aligned grids coupled with obstacle-aware fields to ensure collision-free navigation~\cite{jiang2010continuum}. To improve scalability in dynamic settings, planning-oriented systems often employ heterogeneous representations, i.e., combining navigation meshes with space-time domains, to resolve interactions with moving obstacles~\cite{kapadia2013multi}. While effective for navigation, these pipelines predominantly assume a predefined environment model and require extensive pre-processing. This rigidity limits their adaptability to novel scenes where semantic labels or mesh topology may be unavailable or subject to change.

\paragraph{Perception-Based Navigation.} 
To reduce reliance on global knowledge, an alternative line of work couples agents to the environment through egocentric perception. Synthetic vision techniques~\cite{ondvrej2010synthetic,dutra2017gradient} simulate an agent's field of view to sample visible obstacles, deriving control variables directly from relative motion cues, e.g., time-to-collision. While this allows for more autonomous behavior, the perceptual signals employed are strictly geometric. Consequently, incorporating higher-level semantic behaviors, such as adhering to social norms or interpreting scene context, remains difficult without explicit, manual modeling.

\paragraph{Learning-Based and Data-Driven Methods.} 
The shift toward deep reinforcement learning has enabled richer and more flexible behaviors in cluttered environments, replacing hand-crafted heuristics with learned policies~\cite{lee2018crowd,panayiotou2022ccp,hu2021heterogeneous}. These methods typically encode local geometry via ray-casting or occupancy grids, allowing agents to learn complex collision avoidance and group dynamics. More recently, data-driven formulations have moved away from explicit environment modeling entirely, extracting state and action representations directly from trajectory demonstrations to improve naturalness~\cite{charalambous2023greil}. However, jointly optimizing for geometric feasibility and human-like compliance is non-trivial. Designing reward functions that faithfully reflect social norms without extensive manual labeling remains an open problem.

\paragraph{Language-Conditioned Simulation.} 
Recently, Large Language Models (LLMs) have been adopted to crowd simulation, enabling the generation of diverse behaviors from natural language scripts~\cite{ji2024text}. When conditioned on textual descriptions and semantic maps, these models can synthesize scenario-level variations in motion style and intent. Nevertheless, current approaches often decouple high-level semantic reasoning from low-level control; environmental inputs are typically processed as abstract map representations rather than through an agent's closed-loop perception--reasoning--action mechanism.


\begin{figure*}[t]
  \centering
  \includegraphics[width=\textwidth]{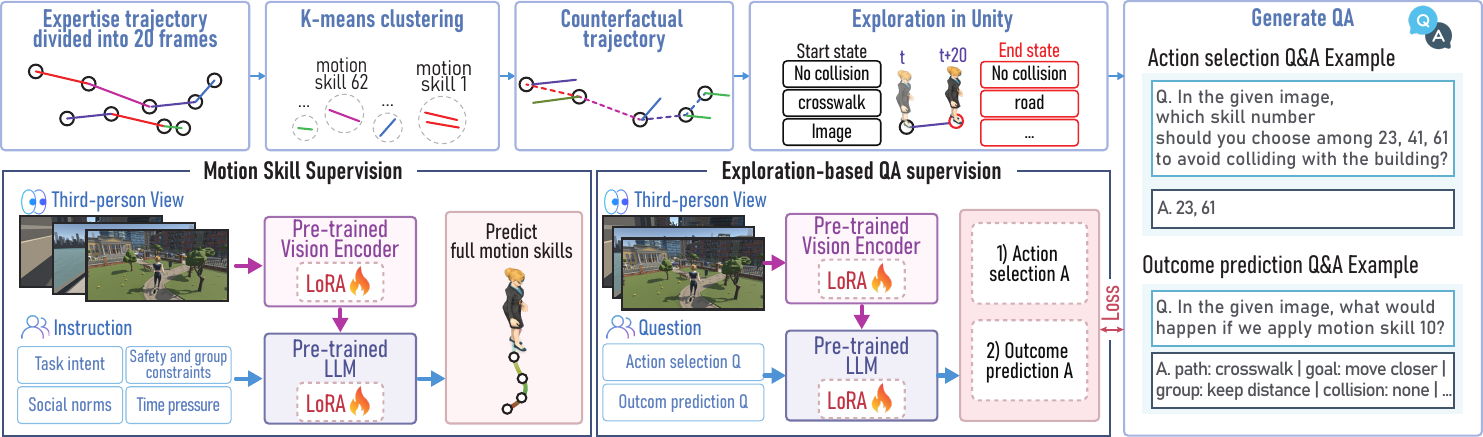}
  \caption{Overview of the dataset processing and training pipeline for Motion Skills and exploration-based QA.
For Motion Skills, long trajectories are segmented into 20-frame clips to construct the skill dataset. For exploration-based QA, segment-level counterfactual trajectories are generated by randomly sampling Motion Skills, and their outcomes are curated into a QA dataset. Using these datasets, a pre-trained vision–language model is fine-tuned with LoRA.}
  \label{fig:Method}
\end{figure*}

\subsection{Vision--Language--Action Policies}
The integration of VLMs into control policies has been proved to be a powerful paradigm for robotics and autonomous systems~\cite{zitkovich2023rt, shao2024lmdrive, durante2025interactive}. By leveraging the extensive world knowledge and semantic alignment inherent in large-scale pre-training, VLA models demonstrate superior generalization to novel objects and instructions compared to policies trained from scratch.

\paragraph{VLA in Robotics and Manipulation.} 
In robotic manipulation, a dominant approach involves fine-tuning Transformer-based backbones (e.g., Llama~\cite{touvron2023llama}) to predict discretized control commands directly from visual and textual inputs~\cite{kim2024openvla, kim2025fine}. These systems prioritize scalability, utilizing massive real-world demonstration datasets to learn robust policies. To facilitate practical deployment, recent works have incorporated parameter-efficient adaptation techniques, such as low-rank adaptation (LoRA) and quantization. However, despite these optimizations, the inherent computational cost of large backbones restricts inference throughput, creating a bottleneck for tasks requiring high-frequency control loops.

\paragraph{Autonomous Driving and Semantic Reasoning.} 
A parallel trend in autonomous driving unifies high-level reasoning and low-level planning within a single autoregressive generator. Built on backbones like Qwen~\cite{bai2025qwen2}, these frameworks tokenize continuous trajectories into physically feasible action tokens, effectively bridging scene understanding and actuation~\cite{zhou2025autovla}. To enhance adaptability, such models often employ dual-process architectures, utilizing direct generation for routine maneuvers and chain-of-thought reasoning for complex scenarios. While effective, the computational latency introduced by long-form reasoning remains a significant barrier to real-time operation.

\paragraph{Efficiency-Oriented Architectures.} 
Recognizing that latency is the primary impediment to VLA deployment, recent research has focused on redesigning supervision and decoding mechanisms for speed. In latency-sensitive 3D environments, methods like CombatVLA~\cite{chen2025combatvla} introduce ``Action-of-Thought'' supervision, progressively training from coarse signals to fine, frame-aligned actions. Furthermore, techniques such as truncated generation, which caps output length via specialized stop tokens, demonstrate that real-time capability relies not just on parameter reduction, but on inference interfaces tailored specifically for fast decision-making.

\paragraph{Gap in Multi-Agent Crowd Simulation.} 
Despite rapid progress in single-agent domains, VLA research in multi-agent or crowd simulation remains limited. Crowd scenarios present unique challenges, for example, interactions are dense, environments are highly dynamic, and the inference cost scales linearly with the number of agents. Our work aims to bridge this gap by bringing the generalization capabilities of VLM-based policies to crowd behavior generation. Building on the end-to-end VLA paradigm, we incorporate efficiency-oriented designs~\cite{kim2024openvla,chen2025combatvla} to ensure that the resulting policies are not only semantically aware and generalizable but also sufficiently responsive for large-scale, real-time multi-agent environments.

\section{CrowdVLA Framework}
Our CrowdVLA Framework is shown in Fig.~\ref{fig:Method}.
Our goal is to enable agents in crowd simulation to infer scene semantics
(e.g., sidewalk/crosswalk/road, obstacles) and social norms
(e.g., avoiding roads, preferring crosswalks, collision avoidance) from the current observation,
and to generate natural, realistic motions without hand-crafted rule-based controllers.
To this end, we learn an end-to-end VLA policy that maps each agent's
third-person view image to an action decision. 


\subsection{Expertise Trajectory Dataset}
Expertise trajectories provide realistic motion priors and stable action grounding. From crowd datasets, we extract pedestrian trajectories, derive motion skills, and construct agent-centric training samples in reconstructed virtual environments running at 25 fps.

\subsubsection{Scene Reconstruction and Agent-centric Observations}
\begin{figure}[t]
  \centering
  \includegraphics[width=1\linewidth]{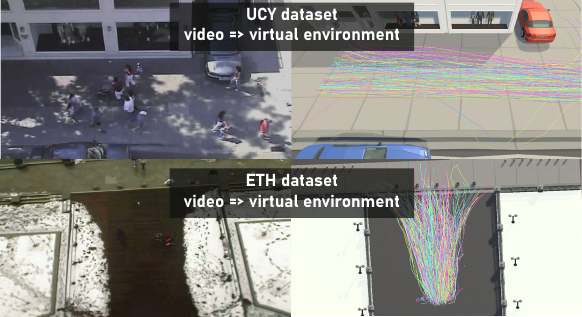}
  \caption{Visualization of Expertise trajectory after transplanting the existing crowd data environment to the unity environment.}
  \label{fig:DataProcessing}
\end{figure}

We use three scenes from the ETH and UCY benchmark~\cite{pellegrini2009you, lerner2007crowds}: Zara2, Univ, and ETH. Each scene is reconstructed in Unity by matching the original layout and geometry, as shown in Fig.~\ref{fig:DataProcessing}. Recorded trajectories are replayed, and a third-person camera is placed slightly behind each agent to capture both the agent and its local surroundings, producing agent-centric observations suitable for VLA training.

\subsubsection{Motion Skill-based Action}
\begin{figure}[t]
  \centering
  \includegraphics[width=1\linewidth]{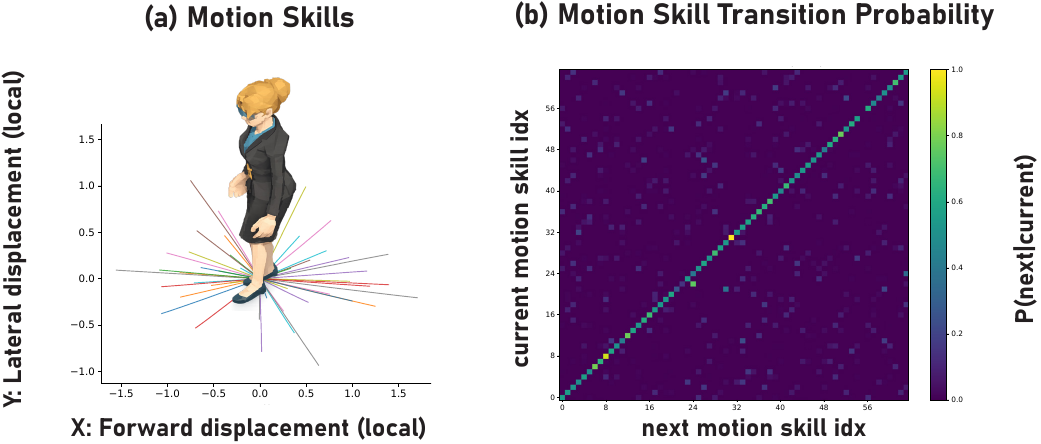}
    \caption{(A) Visualization of representative motion skills as short-horizon trajectory primitives, each encoding a distinct local movement pattern extracted from real crowd data. (B) Empirical transition probabilities between consecutive motion skills estimated from ground-truth trajectories, exhibiting strong diagonal structure that indicates temporal continuity and repetition of similar motion patterns.}
  \label{fig:MotionSkill}
\end{figure}

Instead of regressing continuous control signals, we represent actions as short-horizon future motion segments~\cite{li2025end, chen2024vadv2, li2024hydra}, referred to as \textit{motion skills}. Each motion skill corresponds to a 20-frame trajectory segment extracted from expertise trajectories, as shown in Fig.~\ref{fig:MotionSkill}-(a).

At each decision step, the policy selects one skill conditioned on the current observation, which is executed via a short-horizon tracking controller. This formulation aligns discrete VLM-style action selection with smooth, continuous motion in simulation.

The motion skill set is constructed by clustering 20-frame trajectory segments using K-means; we use $k=64$ skills, which balances expressiveness and stability.
As shown in Fig.~\ref{fig:MotionSkill}-(b), skill transitions exhibit a strong diagonal structure, indicating temporal persistence in which individuals tend to repeat the same or highly similar motion patterns over multiple timesteps~\cite{charalambous2023greil}. This confirms that the motion-skill abstraction preserves the continuity and inertia of pedestrian motion while avoiding abrupt transitions.


\subsubsection{Training Samples from Expertise Trajectories}
Each training sample includes an agent-centric third-person image, the relative positions of the group members, the destination, and remaining time. Remaining time serves as a lightweight conditioning signal that modulates behavior under a single policy. The supervision target is the next motion skill. In total, we collect 15,635 expertise trajectory samples.



\subsection{Exploration-based QA Dataset}
Existing crowd datasets are inherently success-biased: pedestrians usually walk normally, avoid collisions, and follow social norms. As a result, standard supervised learning on demonstrations emphasizes trajectory matching rather than action-conditioned outcome prediction.
This provides little supervision for counterfactual reasoning, leaving models under-specified in anticipating the outcomes of alternative actions.

To address this, we construct an exploration-based QA dataset that explicitly supervises action-conditioned, counterfactual reasoning. Using reconstructed environments, we algorithmically generate alternative motion skills at each timestep that deviate from the demonstrated trajectory and roll them out in the reconstructed environments to observe their outcomes, including collisions, unsafe proximity, norm violations, or inefficient detours.

We formulate two complementary QA types. \textit{Action selection QA} supervises which candidate motion skills satisfy safety or norm constraints under the current observation, allowing multiple valid answers. \textit{Outcome prediction QA} supervises how the environment evolves when executing a specific candidate skill, such as changes in collision risk, goal distance, or group cohesion.

Together, these QA types expose the policy to rare, ambiguous, and non-demonstrated scenarios through counterfactual exploration, enabling it to reason not only about which action to take, but why certain actions lead to safer and more appropriate outcomes.
In total, we collect 359,473 exploration-based QA samples. Additional details are provided in the supplementary material.

\subsection{Training and Inference}
\label{sec:training}
We train a single VLA model using both expertise trajectories and exploration-based QA. The model is initialized from a pretrained VLM and adapted to crowd navigation via LoRA fine-tuning applied to the vision encoder and language model, while freezing all backbone parameters

\subsubsection{Unified Training Data.}
Training is performed in a single stage on a unified dataset that interleaves expertise trajectory samples and exploration-based QA samples, enabling the model to jointly learn motion imitation and action-conditioned reasoning.

\subsubsection{Motion skill Supervision.}
For expertise trajectory samples, the model is trained to predict the demonstrated motion skill.
Given an observation $o_t$, the model outputs a categorical policy $\pi_\theta(\cdot \mid o_t)$ over the motion-skill vocabulary, supervised with cross-entropy:
\begin{equation}
\mathcal{L}_{\text{sup}} = -\mathbb{E}_{(o_t,a_t^{\star})}\left[\log \pi_\theta(a_t^{\star}\mid o_t)\right].
\end{equation}
where $a_t^{\star}$ denotes the ground-truth skill extracted from real crowd trajectories. 

\subsubsection{Exploration-based QA Supervision.}
For QA samples, training focuses on learning action-conditioned outcome reasoning.
Each sample is formatted as a language prompt containing the observation, a question, and candidate motion skills. The supervision target is the corresponding answer, which encodes comparative judgments or predicted outcomes derived from simulator rollouts. 

We train the language model via standard next-token prediction: 
\begin{equation}
\mathcal{L}_{\text{QA}} = -\mathbb{E}_{(Q,A)\sim \mathcal{D}_{\text{QA}}}\left[\log p_\theta(A \mid Q)\right],
\end{equation}
where $Q$ is a language prompt and $A$ is the target answer.

\paragraph{Joint Optimization}
The model is optimized with a unified objective, applying the corresponding loss per sample:
\begin{equation}
\mathcal{L} = \mathcal{L}_{\text{sup}} + \mathcal{L}_{\text{QA}}.
\end{equation}

\paragraph{Multi-horizon Skill Prediction.}
In addition to next-step prediction, the model is trained to predict future motion skill sequences up to episode termination, encouraging temporal coherence and goal-consistent behavior beyond single-step decisions.

\paragraph{Inference.}
At inference time, the model predicts only the next motion skill at each step, executes it, and replans at the following timestep. This minimizes per-step computation and enables real-time performance, while preserving temporal coherence learned during training.

\section{Evaluation}
We evaluate CrowdVLA from both systems and behavioral perspectives, focusing on whether Vision–Language–Action reasoning enables more accurate, socially grounded, and diverse crowd behaviors under identical environmental conditions.

\subsection{Implementation Details}
We used Qwen3-VL-2B-Instruct~\cite{Bai2025Qwen3VLTR} as the backbone VLM and fine-tune it using LoRA on eight NVIDIA A100 Tensor Core GPUs. 
Further Implementation details are provided in the Supplementary Material.

\begin{figure}[t]
  \centering
  \includegraphics[width=1\linewidth]{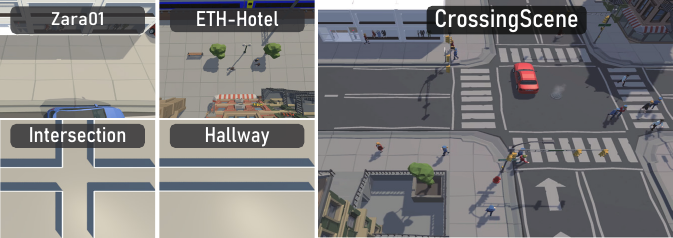}
    \caption{Simulation Environments. Zara01 and ETH-Hotel are reconstructed from the videos provided in the corresponding datasets. Intersection and Hallway are simplified scenes for clearer motion visualization. CrossingScene depicts an everyday-life scenario. }
  \label{fig:SceneExplanation}
\end{figure}

\subsection{Environment Setup}
We evaluate our method across five environments designed to test generalization, robustness, and social norm compliance. Fig. \ref{fig:SceneExplanation} summarizes the evaluation environments.

First, we consider two unseen real-world pedestrian datasets, Zara01 and ETH-Hotel, which are not used during training. These scenes test whether learned behaviors generalize beyond the environments used to construct the expertise trajectories.

Second, we include two standard synthetic scenarios, Hallway and Intersection, which are widely adopted in prior crowd simulation work and primarily stress collision avoidance and goal-directed navigation.

However, many commonly used benchmarks provide limited opportunity to evaluate explicit social norm compliance, as they lack semantically distinct regions with normative constraints. To address this gap, we additionally construct CrossingScene, a daily-life urban scene that explicitly includes both streets and crosswalks, enabling direct comparison of socially normative behaviors such as road avoidance and proper crosswalk usage.


\subsection{Quantitative Comparison}

We compare CrowdVLA against two representative baselines: (i) CCP~\cite{panayiotou2022ccp} as a learning-based RL approach and (ii) GBM~\cite{dutra2017gradient} as a classic synthetic-vision controller.
This choice covers two widely used paradigms for environment-aware crowd navigation.
All methods are evaluated in the same Unity environments with identical initial conditions.

\subsubsection{Metrics}
 \label{subsubsec:Metrics}
We evaluate CrowdVLA along four axes: short-horizon dynamical accuracy, social/semantic compliance, task completion, and behavioral diversity. Full definitions and implementation details are provided in the supplementary.

\paragraph{Progressive Difference Metrics (PDM)}
PDM measures transition accuracy, focusing on whether the policy selects locally correct motion given the current scene and interactions. We evaluate under teacher forcing by resetting the simulator to the ground-truth state at each decision point and assessing only the next predicted segment, avoiding drift accumulation. We report two segment-level variants analogous to ADE and FDE (in meters): PDM-ADE averages the displacement error over all frames in each predicted segment, while PDM-FDE measures the segment's displacement error in final-frame. Lower PDM-ADE/PDM-FDE indicates more accurate local dynamics and better decision quality.

\paragraph{Social Region Violation Rate (SRVR)}
SRVR quantifies norm compliance by measuring how often generated trajectories enter scene-dependent forbidden regions defined by semantic masks and scene conventions. Lower SRVR indicates better adherence to social and semantic constraints.

\paragraph{Goal Success Rate (GSR)}
GSR measures task completion as the fraction of agents that reach their assigned goals within a tolerance before the time limit. Higher GSR indicates more reliable goal-directed behavior.

\paragraph{Trajectory Diversity under Identical Inputs (Div)}
Div is measured by sampling multiple rollouts from identical initial conditions (via different decoding randomness) and computing the average pairwise DTW distance among the resulting trajectories.

\begin{table*}[t]
\centering
\renewcommand{\arraystretch}{1.05}
\caption{Per-scene quantitative results on the held-out test scenes (Zara01 and ETH-Hotel).}
\label{tab:quant_zara01_ethhotel_scene_left}
\begin{tabular*}{\textwidth}{@{\extracolsep{\fill}}@{}llcccc@{}}
\toprule
Scene & Method & PDM-ADE $\downarrow$ & PDM-FDE $\downarrow$ & GSR$\uparrow$ & SRVR $\downarrow$\\
\midrule
\multirow{4}{*}{Zara01}
& GBM & 0.3057$^{\pm .0345}$ & 0.5203$^{\pm .0509}$ & 0.9419$^{\pm .0152}$ & 0.0849$^{\pm .0182}$ \\
& CCP & 0.2406$^{\pm .0005}$ & 0.4524$^{\pm .0005}$ & 0.9797$^{\pm .0000}$ & 0.1796$^{\pm .0018}$ \\
& CCP (Goal Only) & 0.3171$^{\pm .0160}$ & 0.5995$^{\pm .0304}$ & 0.9905$^{\pm .0045}$ & 0.0910$^{\pm .0045}$ \\
& CrowdVLA (ours) & \textbf{0.1288$^{\pm .0031}$} & \textbf{0.2489$^{\pm .0160}$} & \textbf{1.0000$^{\pm .0000}$} & \textbf{0.0593$^{\pm .0104}$} \\
\midrule
\multirow{4}{*}{ETH-Hotel}
& GBM & 0.2262$^{\pm .0104}$ & 0.4306$^{\pm .0198}$ & 0.9821$^{\pm .0072}$ & 0.0749$^{\pm .0082}$ \\
& CCP & 0.2689$^{\pm .0160}$ & 0.4669$^{\pm .0007}$ & 0.7876$^{\pm .0061}$ & 0.1727$^{\pm .0031}$ \\
& CCP (Goal Only) & 0.2694$^{\pm .0221}$ & 0.4816$^{\pm .0433}$ & 0.7758$^{\pm .0359}$ & 0.0958$^{\pm .0051}$ \\
& CrowdVLA (ours) & \textbf{0.2088$^{\pm .0063}$} & \textbf{0.4184$^{\pm .0128}$} & \textbf{0.9909$^{\pm .0038}$}  & \textbf{0.0390$^{\pm .0051}$}  \\
\bottomrule
\end{tabular*}
\end{table*}

\begin{table}[t]
\centering
\setlength{\tabcolsep}{2pt}
\renewcommand{\arraystretch}{1.05}
\caption{Per-scene quantitative results on the held-out test scenes.}
\label{tab:quant_intersection_hallway_scene_left}
\begin{tabular*}{\columnwidth}{@{\extracolsep{\fill}}@{}llcc@{}}
\toprule
Scene & Method & GSR $\uparrow$ & SRVR $\downarrow$\\
\midrule
\multirow{4}{*}{Intersection}
& GBM & 1.0000$^{\pm .0000}$ & 0.1475$^{\pm .1000}$ \\
& CCP & 1.0000$^{\pm .0000}$ & 0.0414$^{\pm .0002}$ \\
& CCP (Goal Only) & 0.9555$^{\pm .0308}$ & 0.0290$^{\pm .0002}$ \\
& CrowdVLA (ours) & \textbf{1.0000$^{\pm .0000}$} & \textbf{0.0170$^{\pm .0158}$} \\
\midrule
\multirow{4}{*}{Hallway}
& GBM & 1.0000$^{\pm .0000}$ & 0.0615$^{\pm .0271}$ \\
& CCP & 1.0000$^{\pm .0000}$ & 0.1585$^{\pm .0000}$ \\
& CCP (Goal Only) & 1.0000$^{\pm .0000}$ & 0.0067$^{\pm .0000}$ \\
& CrowdVLA (ours) & \textbf{1.0000$^{\pm .0000}$} & \textbf{0.0016$^{\pm .0015}$} \\
\bottomrule
\end{tabular*}
\end{table}

\begin{table}[t]
\centering
\caption{Ablation study on held-out test scenes (Zara01 and ETH\_Hotel).}
\label{tab:Ablation}
\begin{tabular*}{\columnwidth}{@{\extracolsep{\fill}}@{}llcc@{}}
\toprule
Scene & Method & PDM-ADE $\downarrow$ & PDM-FDE $\downarrow$\\
\midrule
\multirow{4}{*}{Zara01}
& CrowdVLA (ours) & \textbf{0.1288$^{\pm .0031}$} & \textbf{0.2489$^{\pm .0160}$} \\
& w/o Motion Skill & 0.5876$^{\pm .0035}$ & 1.1714$^{\pm .0055}$\\
& w/o Full trajectory & 0.2337$^{\pm .0024}$ & 0.4861$^{\pm .0018}$\\
& w/o QA & 0.1974$^{\pm .0019}$ & 0.3986$^{\pm .0047}$ \\
\midrule
\multirow{4}{*}{ETH\_Hotel}
& CrowdVLA (ours) & \textbf{0.2088$^{\pm .0063}$} & \textbf{0.4184$^{\pm .0128}$} \\
& w/o Motion Skill & 0.5978$^{\pm .0028}$ & 1.2003$^{\pm .0023}$\\
& w/o Full trajectory & 0.3428$^{\pm .0252}$ & 0.7311$^{\pm .0670}$\\
& w/o QA & 0.2482$^{\pm .0049}$ & 0.5388$^{\pm .0062}$ \\
\bottomrule
\end{tabular*}
\end{table}

\begin{table}[t]
\centering
\caption{Ablation results on held-out test scenes.}
\label{tab:Ablation-Per-Scene}
\begin{tabular*}{\columnwidth}{@{\extracolsep{\fill}}@{}llcc@{}}
\toprule
Scene & Method & GSR $\uparrow$ & SRVR $\downarrow$\\
\midrule
\multirow{4}{*}{Intersection}
& CrowdVLA (ours) & \textbf{1.0000$^{\pm .0000}$} & \textbf{0.0170$^{\pm .0158}$} \\
& w/o Motion Skill & 1.0000$^{\pm .0000}$ & 0.0639$^{\pm .0425}$ \\
& w/o Full trajectory & 1.0000$^{\pm .0000}$ & 0.1124$^{\pm .0445}$ \\
& w/o QA & 0.9444$^{\pm .0845}$ & 0.1369$^{\pm .0537}$ \\
\midrule
\multirow{4}{*}{Hallway}
& CrowdVLA (ours) & \textbf{1.0000$^{\pm .0000}$} & \textbf{0.0016$^{\pm .0015}$} \\
& w/o Motion Skill & 1.0000$^{\pm .0000}$ & 0.0123$^{\pm .0109}$ \\
& w/o Full trajectory & 1.0000$^{\pm .0000}$ & 0.1450$^{\pm .0356}$ \\
& w/o QA & 0.8888$^{\pm .0845}$ & 0.2777$^{\pm .0577}$ \\
\bottomrule
\end{tabular*}
\end{table}

\subsubsection{Overall Results}
Tab.~\ref{tab:quant_zara01_ethhotel_scene_left} and ~\ref{tab:quant_intersection_hallway_scene_left} report results across four axes: short-horizon fidelity (PDM-ADE and PDM-FDE), semantic and social compliance (SRVR) and goal success (GSR).

\paragraph{Short-horizon fidelity (PDM-ADE and PDM-FDE)}
CrowdVLA consistently achieves lower PDM-ADE and PDM-FDE across all scenes, indicating more accurate local motion execution over each motion-skill window. This reflects stable short-horizon decisions under varying scene layouts and interaction contexts. 

\begin{figure*}
  \centering
  \includegraphics[width=1\linewidth]{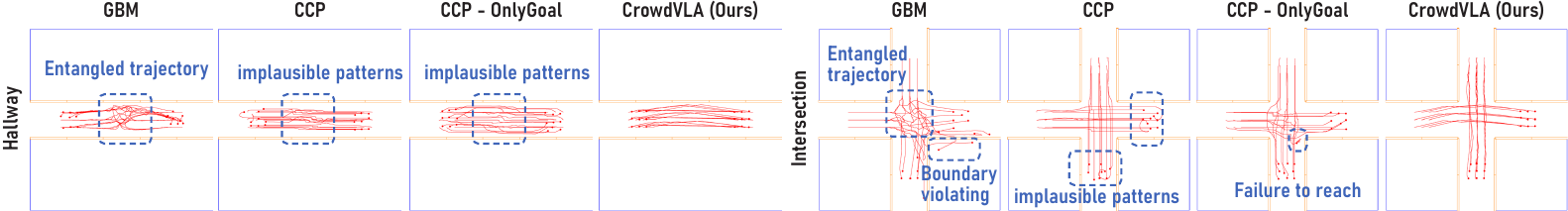}
  \caption{Trajectory Comparison in Hall and Intersection scenes. We compare pedestrian trajectories when nine agents cross in each scene. While CCP~\cite{panayiotou2022ccp} and GBM~\cite{dutra2017gradient} incorporate environmental constraints, their limited integration with expertise trajectory data can result in inconsistent behaviors, including agents failing to reach their goals, becoming entangled in congested regions, or occasionally violating scene boundaries.}
  \label{fig:Comparison_Hall_Intersection}
\end{figure*}

\paragraph{Semantic and social compliance (SRVR) vs.\ task completion (GSR)}
CrowdVLA improves semantic and social norm compliance while maintaining, and often improving, goal success. As shown in Fig. ~\ref{fig:Comparison_Hall_Intersection}, CrowdVLA reaches the goal coherently, while baselines frequently show unstable or implausible behaviors.

Agents reduce violations of socially disallowed regions without sacrificing task completion, avoiding the trade-off commonly observed in baseline methods. This behavior arises from explicitly reasoning about environmental semantics and the consequences of norm-violating actions, rather than relying solely on geometric costs or reward tuning.

\paragraph{Diversity under identical inputs (Div).}
While GBM reports the highest Div, this variability largely stems from unstable behaviors, including agents failing to reach their goals or showing unbounded trajectories, rather than from meaningful alternative strategies. CCP, on the other hand, shows limited diversity under fixed conditioning, as most variation arises from changing profile weights. In contrast, CrowdVLA achieves balanced diversity that coexists with improved SRVR and GSR, demonstrating the emergence of multiple plausible, goal-consistent behaviors rather than failure-driven variance. 
Supplementary material provides detailed quantitative results. 

\subsection{Behavioral Response to Contextual Conditioning}
We visually analyze how contextual conditioning--remaining time and group context--influence agent behavior. For controlled comparison, remaining time is set proportional to distance-to-goal.

\begin{figure*}
  \centering
  \includegraphics[width=1\linewidth]{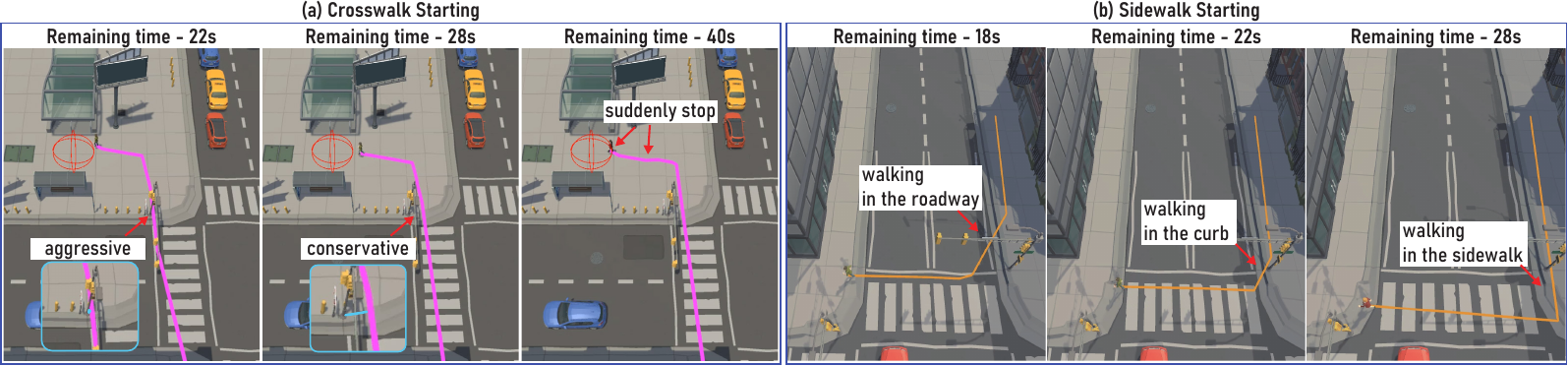}
  \caption{We visualize how trajectories change as remaining time varies: a 1:1 mapping from distance-to-goal, a more urgent setting, and a more relaxed setting.
(a) The agent starts from the crosswalk and moves toward the sidewalk. Under an urgent setting, the trajectory approaches closer to the traffic light (more aggressive), while the default setting stays farther away. Under the relaxed setting, the agent may briefly stop and wait before continuing, which appears to reflect behaviors in the dataset such as pausing unexpectedly to use a phone or check shoes.
(b) The agent starts from the sidewalk with a goal across the crosswalk. Similarly, when time is urgent, the agent partially tolerates the roadway and moves along the crosswalk boundary; with the default setting, it moves toward the curb (the boundary between the crosswalk and sidewalk); with a relaxed setting, it follows a clearly safer trajectory toward the center of the crosswalk. Please refer to the supplementary video for more details.}
  \label{fig:RemainingTime}
\end{figure*}

As shown in .~\ref{fig:RemainingTime}, shorter time budgets lead to more aggressive goal-directed motion and tighter navigation. In urgent cases, agents may temporarily relax certain norms, such as briefly entering the roadway in vehicle-free scenes. This behavior arises from exploration-based QA supervision, which trains the model to distinguish unsafe actions from contextually acceptable deviations, enabling risk-aware decision making rather than rigid rule adherence.

\begin{figure*}
  \centering
  \includegraphics[width=1\linewidth]{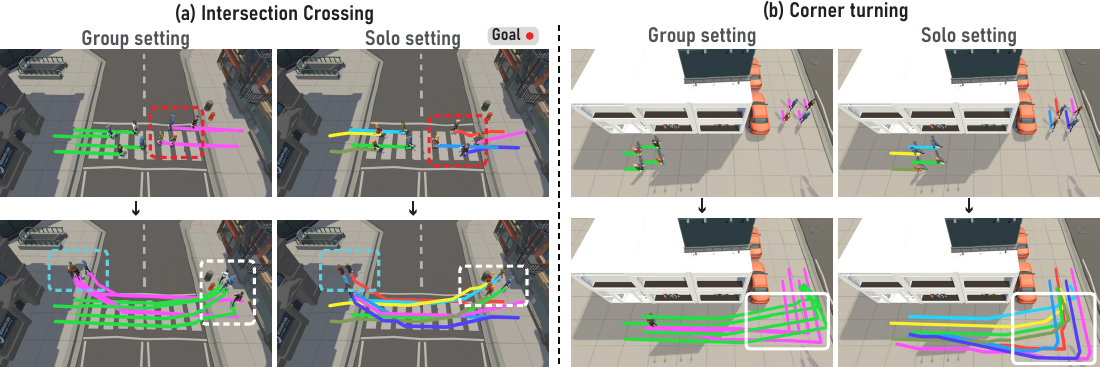}
  \caption{This figure visualizes trajectory differences between group and solo settings under two scenarios: intersection crossing and corner turning.
(a) The red dashed line shows that agents in the solo setting tend to drift farther apart than in the group setting. The blue dashed line shows that group agents move while keeping a closer inter-agent distance. The white dashed line shows that, while the group maintains spacing, solo agents keep distance and pass between obstacles.
(b) The white solid line indicates that, in the solo setting, the agent takes a wider turn around the corner to avoid getting too close to others. Please refer to the supplementary video for more details.}
  \label{fig:GroupingSolo}
\end{figure*}

Fig.~\ref{fig:GroupingSolo} compares solo and group settings. In solo cases, agents increase interpersonal distance while progressing toward the goal. In group settings, agents maintain cohesive spacing while navigating collectively, even in unseen environments.

\subsection{Stress Test}
\begin{figure*}
  \centering
  \includegraphics[width=1\linewidth]{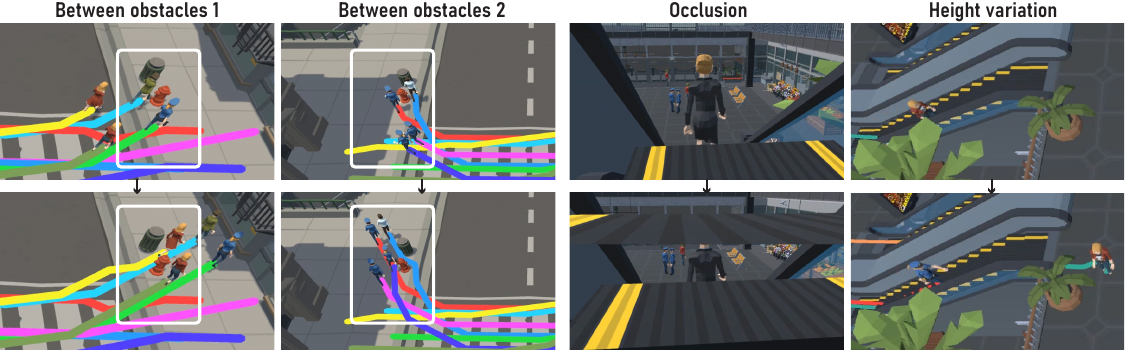}
  \caption{This figure shows robustness tests across multiple unseen environments and scenarios. As in the first and second panels, agents can pass through narrow gaps between obstacles (white solid lines). We also observe that, under occlusion caused by descending an escalator, the model does not panic and still reaches the goal. Conversely, even in unseen cases with height changes (e.g., riding an escalator), the agent adapts robustly. Please refer to the supplementary video for more details.}
  \label{fig:StressTest}
\end{figure*}

To assess the robustness of CrowdVLA under multiple unseen settings. 
As shown in Fig.~\ref{fig:StressTest}, the model remains effective in narrow-gap traversal, occlusion, and height variation scenarios, none of which are observed during training. CrowdVLA relies solely on visual input and requires no additional scene annotations, enabling the same policy to generalize across diverse environments.

\begin{figure*}[t]
  \centering
  \includegraphics[width=1\linewidth]{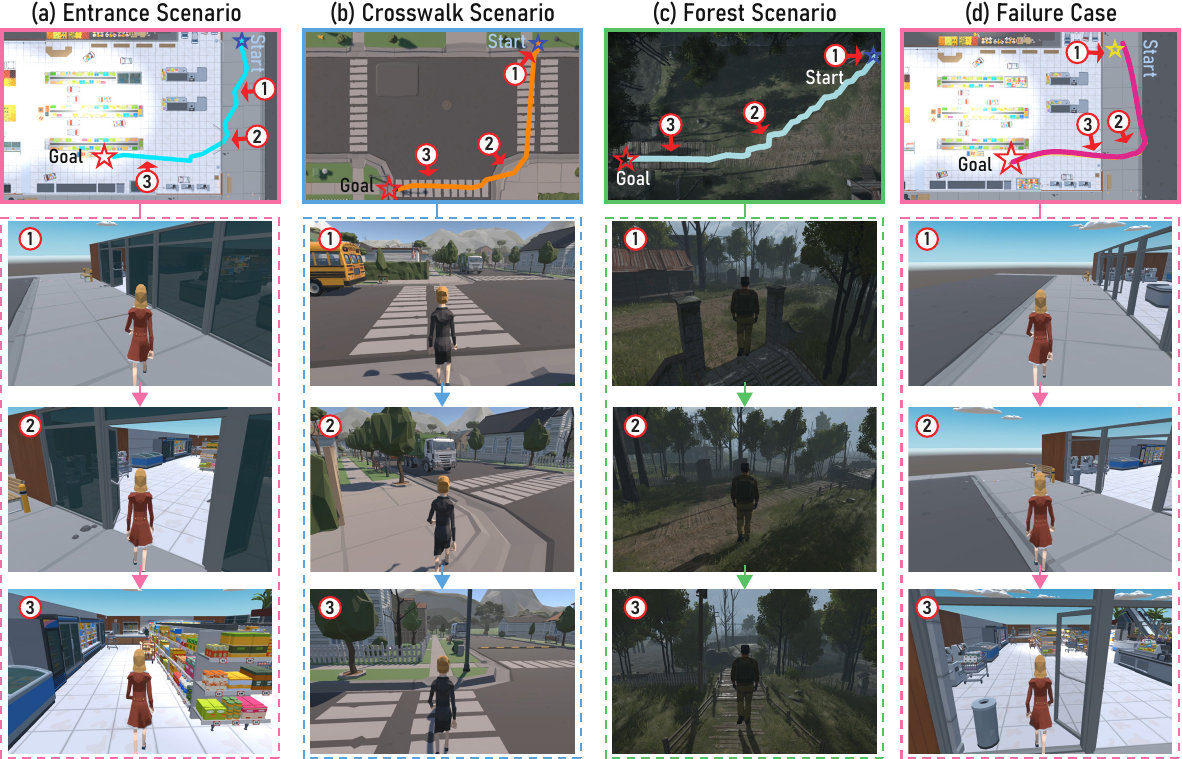}
  \caption{Qualitative stress-test results across unseen environments with similar start–goal configurations. Each column shows the agent's trajectory (top) and corresponding third-person observations at selected timesteps (bottom). Despite the similar start and goal locations, agents adapt their behavior to scene-specific context, including indoor entrances, crosswalks, and natural terrain. The last column illustrates a failure case near transparent glass, where ambiguous visual cues lead to incorrect obstacle perception.}
  \label{fig:StressTest_Single}
\end{figure*}

We further examine agent behavior across environments that share similar start and goal locations but differ in semantics and layout, as shown in Fig.~\ref{fig:StressTest_Single}. 
Agents correctly transition through doorways in entrance scenes, respect crosswalk norms in unseen low-poly environments, and select more direct diagonal paths in dark, realistic forest scenes when unobstructed.
These results indicate consistent goal-directed, collision-aware, and norm-compliant behavior across diverse scene geometries and appearances.

Interestingly, failures near transparent glass arise from perceptual ambiguity rather than control errors. Similar to humans, the agent relies on visual cues and may misinterpret transparent surfaces as traversable space.

\subsection{Ablation Study}
\label{sec:ablation}

We perform an ablation study to quantify the contributions of (i) motion skill-based action representation and (ii) exploration-based QA supervision. All variants share the same Qwen3-VL-2B backbone, LoRA configuration, training schedule, and data splits, and are evaluated under the same simulator settings.

\subsubsection{Variants}
We compare the full model against the following ablations:
(1) w/o Motion Skill: replace discrete motion-skill selection with direct continuous action regression;
(2) w/o QA: train only with expertise trajectories without exploration-based QA;
(3) w/o Full trajectory: disable multi-horizon supervision and predict the only next skill.
Ablations that remove individual QA subtypes are reported in the supplementary material.

\subsubsection{Results and analysis}
The results of the ablation study are presented in Tab.~\ref{tab:Ablation} and ~\ref{tab:Ablation-Per-Scene}.
Across held-out real scenes and synthetic scenarios, CrowdVLA shows consistent improvements. 
Motion skill abstraction stabilizes short-horizon execution by constraining decisions to executable primitives, substantially reducing PDM errors and rollout instability, especially in dense multi-agent interactions where small control noise can cascade. Compared to direct continuous regression, motion skills yield more reliable transitions and enable controlled stochasticity at the decision level.

Exploration-based QA supervision primarily improves semantic and social compliance. Training on success-only demonstrations under-exposes the model to unsafe or norm-violating alternatives, leading to brittle behavior in rare or ambiguous cases. By contrasting counterfactual motion skills and supervising their outcomes, QA explicitly teaches consequence-aware decision making. This sharply reduces SRVR in convention-heavy scenes (e.g., crosswalk vs. roadway) while maintaining or improving goal success, avoiding the typical trade-off between compliance and task completion.

Multi-horizon supervision further improves temporal coherence, reducing myopic switching and stabilizing final segment outcomes (PDM-FDE).

Together, these components address distinct failure modes: motion skills improve dynamical fidelity and controllability, exploration-based QA supplies supervision that is absent from imitation alone, and multi-horizon supervision enforces temporal consistency. Their combination yields lower PDM, lower SRVR, and higher diversity under identical inputs, demonstrating both robust execution with flexible, context-aware behavior.

\subsubsection{Inference Time}
VLM-based policies face scalability challenges as agent count grows. CrowdVLA mitigates this by predicting actions at a 20-frame motion-skill cadence, using lightweight LoRA-based adaptation, and batching agent queries for efficient GPU utilization. Compared to per-frame control prediction, the motion-skill interface avoids repeated inference overhead and reduces end-to-end rollout time for a fixed horizon. While more than thousands of agents remain compute-bound, this design substantially improves practical deployability and points toward future system-level optimizations such as quantization and asynchronous serving. During real-time inference, to ensure smooth, uninterrupted motion, we ran the model several steps ahead; details are provided in the supplementary material.

\section{Conclusion}
\label{sec:conclusion}
CrowdVLA reframes crowd simulation as a problem of perception-driven, consequence-aware decision making, rather than rule execution or trajectory replay. By grounding individual agents in a Vision--Language--Action loop, our approach enables pedestrians to interpret scene semantics and social context directly from visual observations, and to select actions based on their consequences.

This formulation is made practical through three key design choices. A pretrained VLM is adapted to pedestrian control via LoRA fine-tuning, preserving broad visual and commonsense priors while remaining computationally feasible. Action selection is mediated through a motion-skill vocabulary, where each token represents a short, executable trajectory segment, bridging token-based reasoning and continuous simulation. Finally, exploration-based supervision exposes the policy to counterfactual alternatives and their consequences, overcoming the strong success bias of recorded crowd data and enabling risk- and norm-aware decisions.

Several limitations remain. Motion-skill discretization may constrain fine-grained control. Norm compliance currently relies on scene-specific semantic definitions, and longer-range intent, such as activities or group formation, is only partially modeled. Addressing these challenges will require richer scene understanding, hierarchical action abstractions, and extended temporal reasoning.

This work suggests a shift in crowd simulation from motion-centric synthesis to agents that see, reason, and act within their environment. By casting VLA reasoning as a unifying interface between perception, social understanding, and motion, CrowdVLA enables crowd behaviors that are semantically grounded, adaptable to new scenes, and driven by understanding rather than predefined rules or demonstration-driven imitation.

\bibliographystyle{ACM-Reference-Format}
\bibliography{references}


\begin{thebibliography}{36}


\ifx \showCODEN    \undefined \def \showCODEN     #1{\unskip}     \fi
\ifx \showISBNx    \undefined \def \showISBNx     #1{\unskip}     \fi
\ifx \showISBNxiii \undefined \def \showISBNxiii  #1{\unskip}     \fi
\ifx \showISSN     \undefined \def \showISSN      #1{\unskip}     \fi
\ifx \showLCCN     \undefined \def \showLCCN      #1{\unskip}     \fi
\ifx \shownote     \undefined \def \shownote      #1{#1}          \fi
\ifx \showarticletitle \undefined \def \showarticletitle #1{#1}   \fi
\ifx \showURL      \undefined \def \showURL       {\relax}        \fi
\providecommand\bibfield[2]{#2}
\providecommand\bibinfo[2]{#2}
\providecommand\natexlab[1]{#1}
\providecommand\showeprint[2][]{arXiv:#2}

\bibitem[Bai et~al\mbox{.}(2025a)]%
        {Bai2025Qwen3VLTR}
\bibfield{author}{\bibinfo{person}{Shuai Bai}, \bibinfo{person}{Yuxuan Cai}, \bibinfo{person}{Ruizhe Chen}, \bibinfo{person}{Keqin Chen}, \bibinfo{person}{Xiong-Hui Chen}, \bibinfo{person}{Zesen Cheng}, \bibinfo{person}{Lianghao Deng}, \bibinfo{person}{Wei Ding}, \bibinfo{person}{Rongyao Fang}, \bibinfo{person}{Chang Gao}, \bibinfo{person}{Chunjiang Ge}, \bibinfo{person}{Wenbin Ge}, \bibinfo{person}{Zhifang Guo}, \bibinfo{person}{Qidong Huang}, \bibinfo{person}{Jie Huang}, \bibinfo{person}{Fei Huang}, \bibinfo{person}{Binyuan Hui}, \bibinfo{person}{Shutong Jiang}, \bibinfo{person}{Zhaohai Li}, \bibinfo{person}{Mingsheng Li}, \bibinfo{person}{Mei Li}, \bibinfo{person}{Kaixin Li}, \bibinfo{person}{Zicheng Lin}, \bibinfo{person}{Junyang Lin}, \bibinfo{person}{Xuejing Liu}, \bibinfo{person}{Jiawei Liu}, \bibinfo{person}{Chenglong Liu}, \bibinfo{person}{Yang Liu}, \bibinfo{person}{Dayiheng Liu}, \bibinfo{person}{Shixuan Liu}, \bibinfo{person}{Dunjie Lu}, \bibinfo{person}{Ruilin Luo}, \bibinfo{person}{Chenxu Lv},
  \bibinfo{person}{Rui Men}, \bibinfo{person}{Li~Ying Meng}, \bibinfo{person}{Xuancheng Ren}, \bibinfo{person}{Xin yi Ren}, \bibinfo{person}{Sibo Song}, \bibinfo{person}{Yu-Chen Sun}, \bibinfo{person}{Jun Tang}, \bibinfo{person}{Jianhong Tu}, \bibinfo{person}{Jianqiang Wan}, \bibinfo{person}{Peng Wang}, \bibinfo{person}{Pengfei Wang}, \bibinfo{person}{Qiuyue Wang}, \bibinfo{person}{Yuxuan Wang}, \bibinfo{person}{Tianbao Xie}, \bibinfo{person}{Yihe Xu}, \bibinfo{person}{Haiyang Xu}, \bibinfo{person}{Jin Xu}, \bibinfo{person}{Zhibo Yang}, \bibinfo{person}{Mingkun Yang}, \bibinfo{person}{Jian-Xing Yang}, \bibinfo{person}{An Yang}, \bibinfo{person}{Bowen Yu}, \bibinfo{person}{Fei Zhang}, \bibinfo{person}{Hang Zhang}, \bibinfo{person}{Xi Zhang}, \bibinfo{person}{Botao Zheng}, \bibinfo{person}{Humen Zhong}, \bibinfo{person}{Jingren Zhou}, \bibinfo{person}{Fanxi Zhou}, \bibinfo{person}{Jingren Zhou}, \bibinfo{person}{Yuanzhi Zhu}, {and} \bibinfo{person}{Keming Zhu}.} \bibinfo{year}{2025}\natexlab{a}.
\newblock \showarticletitle{Qwen3-VL Technical Report}.
\newblock
\urldef\tempurl%
\url{https://api.semanticscholar.org/CorpusID:283262018}
\showURL{%
\tempurl}


\bibitem[Bai et~al\mbox{.}(2025b)]%
        {bai2025qwen2}
\bibfield{author}{\bibinfo{person}{Shuai Bai}, \bibinfo{person}{Keqin Chen}, \bibinfo{person}{Xuejing Liu}, \bibinfo{person}{Jialin Wang}, \bibinfo{person}{Wenbin Ge}, \bibinfo{person}{Sibo Song}, \bibinfo{person}{Kai Dang}, \bibinfo{person}{Peng Wang}, \bibinfo{person}{Shijie Wang}, \bibinfo{person}{Jun Tang}, {et~al\mbox{.}}} \bibinfo{year}{2025}\natexlab{b}.
\newblock \showarticletitle{Qwen2. 5-vl technical report}.
\newblock \bibinfo{journal}{\emph{arXiv preprint arXiv:2502.13923}} (\bibinfo{year}{2025}).
\newblock


\bibitem[Charalambous et~al\mbox{.}(2023)]%
        {charalambous2023greil}
\bibfield{author}{\bibinfo{person}{Panayiotis Charalambous}, \bibinfo{person}{Julien Pettre}, \bibinfo{person}{Vassilis Vassiliades}, \bibinfo{person}{Yiorgos Chrysanthou}, {and} \bibinfo{person}{Nuria Pelechano}.} \bibinfo{year}{2023}\natexlab{}.
\newblock \showarticletitle{Greil-crowds: Crowd simulation with deep reinforcement learning and examples}.
\newblock \bibinfo{journal}{\emph{ACM Transactions on Graphics (TOG)}} \bibinfo{volume}{42}, \bibinfo{number}{4} (\bibinfo{year}{2023}), \bibinfo{pages}{1--15}.
\newblock


\bibitem[Chen et~al\mbox{.}(2025)]%
        {chen2025combatvla}
\bibfield{author}{\bibinfo{person}{Peng Chen}, \bibinfo{person}{Pi Bu}, \bibinfo{person}{Yingyao Wang}, \bibinfo{person}{Xinyi Wang}, \bibinfo{person}{Ziming Wang}, \bibinfo{person}{Jie Guo}, \bibinfo{person}{Yingxiu Zhao}, \bibinfo{person}{Qi Zhu}, \bibinfo{person}{Jun Song}, \bibinfo{person}{Siran Yang}, {et~al\mbox{.}}} \bibinfo{year}{2025}\natexlab{}.
\newblock \showarticletitle{Combatvla: An efficient vision-language-action model for combat tasks in 3d action role-playing games}.
\newblock \bibinfo{journal}{\emph{arXiv preprint arXiv:2503.09527}} (\bibinfo{year}{2025}).
\newblock


\bibitem[Chen et~al\mbox{.}(2024)]%
        {chen2024vadv2}
\bibfield{author}{\bibinfo{person}{Shaoyu Chen}, \bibinfo{person}{Bo Jiang}, \bibinfo{person}{Hao Gao}, \bibinfo{person}{Bencheng Liao}, \bibinfo{person}{Qing Xu}, \bibinfo{person}{Qian Zhang}, \bibinfo{person}{Chang Huang}, \bibinfo{person}{Wenyu Liu}, {and} \bibinfo{person}{Xinggang Wang}.} \bibinfo{year}{2024}\natexlab{}.
\newblock \showarticletitle{Vadv2: End-to-end vectorized autonomous driving via probabilistic planning}.
\newblock \bibinfo{journal}{\emph{arXiv preprint arXiv:2402.13243}} (\bibinfo{year}{2024}).
\newblock


\bibitem[Durante et~al\mbox{.}(2025)]%
        {durante2025interactive}
\bibfield{author}{\bibinfo{person}{Zane Durante}, \bibinfo{person}{Ran Gong}, \bibinfo{person}{Bidipta Sarkar}, \bibinfo{person}{Naoki Wake}, \bibinfo{person}{Rohan Taori}, \bibinfo{person}{Paul Tang}, \bibinfo{person}{Shrinidhi Lakshmikanth}, \bibinfo{person}{Kevin Schulman}, \bibinfo{person}{Arnold Milstein}, \bibinfo{person}{Hoi Vo}, {et~al\mbox{.}}} \bibinfo{year}{2025}\natexlab{}.
\newblock \showarticletitle{An interactive agent foundation model}. In \bibinfo{booktitle}{\emph{Proceedings of the Computer Vision and Pattern Recognition Conference}}. \bibinfo{pages}{3652--3662}.
\newblock


\bibitem[Dutra et~al\mbox{.}(2017)]%
        {dutra2017gradient}
\bibfield{author}{\bibinfo{person}{Te{\'o}filo~Bezerra Dutra}, \bibinfo{person}{Ricardo Marques}, \bibinfo{person}{Joaquim~B Cavalcante-Neto}, \bibinfo{person}{Creto~Augusto Vidal}, {and} \bibinfo{person}{Julien Pettr{\'e}}.} \bibinfo{year}{2017}\natexlab{}.
\newblock \showarticletitle{Gradient-based steering for vision-based crowd simulation algorithms}. In \bibinfo{booktitle}{\emph{Computer graphics forum}}, Vol.~\bibinfo{volume}{36}. Wiley Online Library, \bibinfo{pages}{337--348}.
\newblock


\bibitem[Guy et~al\mbox{.}(2010)]%
        {guy2010pledestrians}
\bibfield{author}{\bibinfo{person}{Stephen~J Guy}, \bibinfo{person}{Jatin Chhugani}, \bibinfo{person}{Sean Curtis}, \bibinfo{person}{Pradeep Dubey}, \bibinfo{person}{Ming~C Lin}, {and} \bibinfo{person}{Dinesh Manocha}.} \bibinfo{year}{2010}\natexlab{}.
\newblock \showarticletitle{PLEdestrians: A Least-Effort Approach to Crowd Simulation.}. In \bibinfo{booktitle}{\emph{Symposium on computer animation}}. \bibinfo{pages}{119--128}.
\newblock


\bibitem[Guy et~al\mbox{.}(2009)]%
        {guy2009clearpath}
\bibfield{author}{\bibinfo{person}{Stephen~J Guy}, \bibinfo{person}{Jatin Chhugani}, \bibinfo{person}{Changkyu Kim}, \bibinfo{person}{Nadathur Satish}, \bibinfo{person}{Ming Lin}, \bibinfo{person}{Dinesh Manocha}, {and} \bibinfo{person}{Pradeep Dubey}.} \bibinfo{year}{2009}\natexlab{}.
\newblock \showarticletitle{Clearpath: highly parallel collision avoidance for multi-agent simulation}. In \bibinfo{booktitle}{\emph{Proceedings of the 2009 ACM SIGGRAPH/Eurographics Symposium on Computer Animation}}. \bibinfo{pages}{177--187}.
\newblock


\bibitem[Guy et~al\mbox{.}(2011)]%
        {guy2011simulating}
\bibfield{author}{\bibinfo{person}{Stephen~J Guy}, \bibinfo{person}{Sujeong Kim}, \bibinfo{person}{Ming~C Lin}, {and} \bibinfo{person}{Dinesh Manocha}.} \bibinfo{year}{2011}\natexlab{}.
\newblock \showarticletitle{Simulating heterogeneous crowd behaviors using personality trait theory}. In \bibinfo{booktitle}{\emph{Proceedings of the 2011 ACM SIGGRAPH/Eurographics symposium on computer animation}}. \bibinfo{pages}{43--52}.
\newblock


\bibitem[Helbing and Molnar(1995)]%
        {helbing1995social}
\bibfield{author}{\bibinfo{person}{Dirk Helbing} {and} \bibinfo{person}{Peter Molnar}.} \bibinfo{year}{1995}\natexlab{}.
\newblock \showarticletitle{Social force model for pedestrian dynamics}.
\newblock \bibinfo{journal}{\emph{Physical review E}} \bibinfo{volume}{51}, \bibinfo{number}{5} (\bibinfo{year}{1995}), \bibinfo{pages}{4282}.
\newblock


\bibitem[Hu et~al\mbox{.}(2022)]%
        {hu2022lora}
\bibfield{author}{\bibinfo{person}{Edward~J Hu}, \bibinfo{person}{Yelong Shen}, \bibinfo{person}{Phillip Wallis}, \bibinfo{person}{Zeyuan Allen-Zhu}, \bibinfo{person}{Yuanzhi Li}, \bibinfo{person}{Shean Wang}, \bibinfo{person}{Lu Wang}, \bibinfo{person}{Weizhu Chen}, {et~al\mbox{.}}} \bibinfo{year}{2022}\natexlab{}.
\newblock \showarticletitle{Lora: Low-rank adaptation of large language models.}
\newblock \bibinfo{journal}{\emph{ICLR}} \bibinfo{volume}{1}, \bibinfo{number}{2} (\bibinfo{year}{2022}), \bibinfo{pages}{3}.
\newblock


\bibitem[Hu et~al\mbox{.}(2021)]%
        {hu2021heterogeneous}
\bibfield{author}{\bibinfo{person}{Kaidong Hu}, \bibinfo{person}{Brandon Haworth}, \bibinfo{person}{Glen Berseth}, \bibinfo{person}{Vladimir Pavlovic}, \bibinfo{person}{Petros Faloutsos}, {and} \bibinfo{person}{Mubbasir Kapadia}.} \bibinfo{year}{2021}\natexlab{}.
\newblock \showarticletitle{Heterogeneous crowd simulation using parametric reinforcement learning}.
\newblock \bibinfo{journal}{\emph{IEEE Transactions on Visualization and Computer Graphics}} \bibinfo{volume}{29}, \bibinfo{number}{4} (\bibinfo{year}{2021}), \bibinfo{pages}{2036--2052}.
\newblock


\bibitem[Ji et~al\mbox{.}(2024)]%
        {ji2024text}
\bibfield{author}{\bibinfo{person}{Xuebo Ji}, \bibinfo{person}{Zherong Pan}, \bibinfo{person}{Xifeng Gao}, {and} \bibinfo{person}{Jia Pan}.} \bibinfo{year}{2024}\natexlab{}.
\newblock \showarticletitle{Text-guided synthesis of crowd animation}. In \bibinfo{booktitle}{\emph{ACM SIGGRAPH 2024 Conference Papers}}. \bibinfo{pages}{1--11}.
\newblock


\bibitem[Jiang et~al\mbox{.}(2010)]%
        {jiang2010continuum}
\bibfield{author}{\bibinfo{person}{Hao Jiang}, \bibinfo{person}{Wenbin Xu}, \bibinfo{person}{Tianlu Mao}, \bibinfo{person}{Chunpeng Li}, \bibinfo{person}{Shihong Xia}, {and} \bibinfo{person}{Zhaoqi Wang}.} \bibinfo{year}{2010}\natexlab{}.
\newblock \showarticletitle{Continuum crowd simulation in complex environments}.
\newblock \bibinfo{journal}{\emph{Computers \& Graphics}} \bibinfo{volume}{34}, \bibinfo{number}{5} (\bibinfo{year}{2010}), \bibinfo{pages}{537--544}.
\newblock


\bibitem[Kapadia et~al\mbox{.}(2013)]%
        {kapadia2013multi}
\bibfield{author}{\bibinfo{person}{Mubbasir Kapadia}, \bibinfo{person}{Alejandro Beacco}, \bibinfo{person}{Francisco Garcia}, \bibinfo{person}{Vivek Reddy}, \bibinfo{person}{Nuria Pelechano}, {and} \bibinfo{person}{Norman~I Badler}.} \bibinfo{year}{2013}\natexlab{}.
\newblock \showarticletitle{Multi-domain real-time planning in dynamic environments}. In \bibinfo{booktitle}{\emph{Proceedings of the 12th ACM SIGGRAPH/Eurographics symposium on computer animation}}. \bibinfo{pages}{115--124}.
\newblock


\bibitem[Kim et~al\mbox{.}(2025)]%
        {kim2025fine}
\bibfield{author}{\bibinfo{person}{Moo~Jin Kim}, \bibinfo{person}{Chelsea Finn}, {and} \bibinfo{person}{Percy Liang}.} \bibinfo{year}{2025}\natexlab{}.
\newblock \showarticletitle{Fine-tuning vision-language-action models: Optimizing speed and success}.
\newblock \bibinfo{journal}{\emph{arXiv preprint arXiv:2502.19645}} (\bibinfo{year}{2025}).
\newblock


\bibitem[Kim et~al\mbox{.}(2024)]%
        {kim2024openvla}
\bibfield{author}{\bibinfo{person}{Moo~Jin Kim}, \bibinfo{person}{Karl Pertsch}, \bibinfo{person}{Siddharth Karamcheti}, \bibinfo{person}{Ted Xiao}, \bibinfo{person}{Ashwin Balakrishna}, \bibinfo{person}{Suraj Nair}, \bibinfo{person}{Rafael Rafailov}, \bibinfo{person}{Ethan Foster}, \bibinfo{person}{Grace Lam}, \bibinfo{person}{Pannag Sanketi}, {et~al\mbox{.}}} \bibinfo{year}{2024}\natexlab{}.
\newblock \showarticletitle{Openvla: An open-source vision-language-action model}.
\newblock \bibinfo{journal}{\emph{arXiv preprint arXiv:2406.09246}} (\bibinfo{year}{2024}).
\newblock


\bibitem[Lee et~al\mbox{.}(2018)]%
        {lee2018crowd}
\bibfield{author}{\bibinfo{person}{Jaedong Lee}, \bibinfo{person}{Jungdam Won}, {and} \bibinfo{person}{Jehee Lee}.} \bibinfo{year}{2018}\natexlab{}.
\newblock \showarticletitle{Crowd simulation by deep reinforcement learning}. In \bibinfo{booktitle}{\emph{Proceedings of the 11th ACM SIGGRAPH conference on motion, interaction and games}}. \bibinfo{pages}{1--7}.
\newblock


\bibitem[Lee et~al\mbox{.}(2007)]%
        {lee2007group}
\bibfield{author}{\bibinfo{person}{Kang~Hoon Lee}, \bibinfo{person}{Myung~Geol Choi}, \bibinfo{person}{Qyoun Hong}, {and} \bibinfo{person}{Jehee Lee}.} \bibinfo{year}{2007}\natexlab{}.
\newblock \showarticletitle{Group behavior from video: a data-driven approach to crowd simulation}. In \bibinfo{booktitle}{\emph{Proceedings of the 2007 ACM SIGGRAPH/Eurographics symposium on Computer animation}}. \bibinfo{pages}{109--118}.
\newblock


\bibitem[Lerner et~al\mbox{.}(2007)]%
        {lerner2007crowds}
\bibfield{author}{\bibinfo{person}{Alon Lerner}, \bibinfo{person}{Yiorgos Chrysanthou}, {and} \bibinfo{person}{Dani Lischinski}.} \bibinfo{year}{2007}\natexlab{}.
\newblock \showarticletitle{Crowds by example}. In \bibinfo{booktitle}{\emph{Computer graphics forum}}, Vol.~\bibinfo{volume}{26}. Wiley Online Library, \bibinfo{pages}{655--664}.
\newblock


\bibitem[Li et~al\mbox{.}(2025)]%
        {li2025end}
\bibfield{author}{\bibinfo{person}{Yingyan Li}, \bibinfo{person}{Yuqi Wang}, \bibinfo{person}{Yang Liu}, \bibinfo{person}{Jiawei He}, \bibinfo{person}{Lue Fan}, {and} \bibinfo{person}{Zhaoxiang Zhang}.} \bibinfo{year}{2025}\natexlab{}.
\newblock \showarticletitle{End-to-end driving with online trajectory evaluation via bev world model}.
\newblock \bibinfo{journal}{\emph{arXiv preprint arXiv:2504.01941}} (\bibinfo{year}{2025}).
\newblock


\bibitem[Li et~al\mbox{.}(2024)]%
        {li2024hydra}
\bibfield{author}{\bibinfo{person}{Zhenxin Li}, \bibinfo{person}{Kailin Li}, \bibinfo{person}{Shihao Wang}, \bibinfo{person}{Shiyi Lan}, \bibinfo{person}{Zhiding Yu}, \bibinfo{person}{Yishen Ji}, \bibinfo{person}{Zhiqi Li}, \bibinfo{person}{Ziyue Zhu}, \bibinfo{person}{Jan Kautz}, \bibinfo{person}{Zuxuan Wu}, {et~al\mbox{.}}} \bibinfo{year}{2024}\natexlab{}.
\newblock \showarticletitle{Hydra-mdp: End-to-end multimodal planning with multi-target hydra-distillation}.
\newblock \bibinfo{journal}{\emph{arXiv preprint arXiv:2406.06978}} (\bibinfo{year}{2024}).
\newblock


\bibitem[Ond{\v{r}}ej et~al\mbox{.}(2010)]%
        {ondvrej2010synthetic}
\bibfield{author}{\bibinfo{person}{Jan Ond{\v{r}}ej}, \bibinfo{person}{Julien Pettr{\'e}}, \bibinfo{person}{Anne-H{\'e}l{\`e}ne Olivier}, {and} \bibinfo{person}{St{\'e}phane Donikian}.} \bibinfo{year}{2010}\natexlab{}.
\newblock \showarticletitle{A synthetic-vision based steering approach for crowd simulation}.
\newblock \bibinfo{journal}{\emph{ACM Transactions on Graphics (TOG)}} \bibinfo{volume}{29}, \bibinfo{number}{4} (\bibinfo{year}{2010}), \bibinfo{pages}{1--9}.
\newblock


\bibitem[Panayiotou et~al\mbox{.}(2022)]%
        {panayiotou2022ccp}
\bibfield{author}{\bibinfo{person}{Andreas Panayiotou}, \bibinfo{person}{Theodoros Kyriakou}, \bibinfo{person}{Marilena Lemonari}, \bibinfo{person}{Yiorgos Chrysanthou}, {and} \bibinfo{person}{Panayiotis Charalambous}.} \bibinfo{year}{2022}\natexlab{}.
\newblock \showarticletitle{Ccp: Configurable crowd profiles}. In \bibinfo{booktitle}{\emph{ACM SIGGRAPH 2022 conference proceedings}}. \bibinfo{pages}{1--10}.
\newblock


\bibitem[Pellegrini et~al\mbox{.}(2009)]%
        {pellegrini2009you}
\bibfield{author}{\bibinfo{person}{Stefano Pellegrini}, \bibinfo{person}{Andreas Ess}, \bibinfo{person}{Konrad Schindler}, {and} \bibinfo{person}{Luc Van~Gool}.} \bibinfo{year}{2009}\natexlab{}.
\newblock \showarticletitle{You'll never walk alone: Modeling social behavior for multi-target tracking}. In \bibinfo{booktitle}{\emph{2009 IEEE 12th international conference on computer vision}}. IEEE, \bibinfo{pages}{261--268}.
\newblock


\bibitem[Qiu and Hu(2010)]%
        {qiu2010modeling}
\bibfield{author}{\bibinfo{person}{Fasheng Qiu} {and} \bibinfo{person}{Xiaolin Hu}.} \bibinfo{year}{2010}\natexlab{}.
\newblock \showarticletitle{Modeling group structures in pedestrian crowd simulation}.
\newblock \bibinfo{journal}{\emph{Simulation Modelling Practice and Theory}} \bibinfo{volume}{18}, \bibinfo{number}{2} (\bibinfo{year}{2010}), \bibinfo{pages}{190--205}.
\newblock


\bibitem[Robicquet et~al\mbox{.}(2016)]%
        {robicquet2016learning}
\bibfield{author}{\bibinfo{person}{Alexandre Robicquet}, \bibinfo{person}{Amir Sadeghian}, \bibinfo{person}{Alexandre Alahi}, {and} \bibinfo{person}{Silvio Savarese}.} \bibinfo{year}{2016}\natexlab{}.
\newblock \showarticletitle{Learning social etiquette: Human trajectory understanding in crowded scenes}. In \bibinfo{booktitle}{\emph{European conference on computer vision}}. Springer, \bibinfo{pages}{549--565}.
\newblock


\bibitem[Shao et~al\mbox{.}(2024)]%
        {shao2024lmdrive}
\bibfield{author}{\bibinfo{person}{Hao Shao}, \bibinfo{person}{Yuxuan Hu}, \bibinfo{person}{Letian Wang}, \bibinfo{person}{Guanglu Song}, \bibinfo{person}{Steven~L Waslander}, \bibinfo{person}{Yu Liu}, {and} \bibinfo{person}{Hongsheng Li}.} \bibinfo{year}{2024}\natexlab{}.
\newblock \showarticletitle{Lmdrive: Closed-loop end-to-end driving with large language models}. In \bibinfo{booktitle}{\emph{Proceedings of the IEEE/CVF Conference on Computer Vision and Pattern Recognition}}. \bibinfo{pages}{15120--15130}.
\newblock


\bibitem[Touvron et~al\mbox{.}(2023)]%
        {touvron2023llama}
\bibfield{author}{\bibinfo{person}{Hugo Touvron}, \bibinfo{person}{Louis Martin}, \bibinfo{person}{Kevin Stone}, \bibinfo{person}{Peter Albert}, \bibinfo{person}{Amjad Almahairi}, \bibinfo{person}{Yasmine Babaei}, \bibinfo{person}{Nikolay Bashlykov}, \bibinfo{person}{Soumya Batra}, \bibinfo{person}{Prajjwal Bhargava}, \bibinfo{person}{Shruti Bhosale}, {et~al\mbox{.}}} \bibinfo{year}{2023}\natexlab{}.
\newblock \showarticletitle{Llama 2: Open foundation and fine-tuned chat models}.
\newblock \bibinfo{journal}{\emph{arXiv preprint arXiv:2307.09288}} (\bibinfo{year}{2023}).
\newblock


\bibitem[Van~den Berg et~al\mbox{.}(2008)]%
        {van2008reciprocal}
\bibfield{author}{\bibinfo{person}{Jur Van~den Berg}, \bibinfo{person}{Ming Lin}, {and} \bibinfo{person}{Dinesh Manocha}.} \bibinfo{year}{2008}\natexlab{}.
\newblock \showarticletitle{Reciprocal velocity obstacles for real-time multi-agent navigation}. In \bibinfo{booktitle}{\emph{2008 IEEE international conference on robotics and automation}}. Ieee, \bibinfo{pages}{1928--1935}.
\newblock


\bibitem[Wei et~al\mbox{.}(2018)]%
        {wei2018learning}
\bibfield{author}{\bibinfo{person}{Xiang Wei}, \bibinfo{person}{Wei Lu}, \bibinfo{person}{Lili Zhu}, {and} \bibinfo{person}{Weiwei Xing}.} \bibinfo{year}{2018}\natexlab{}.
\newblock \showarticletitle{Learning motion rules from real data: Neural network for crowd simulation}.
\newblock \bibinfo{journal}{\emph{Neurocomputing}}  \bibinfo{volume}{310} (\bibinfo{year}{2018}), \bibinfo{pages}{125--134}.
\newblock


\bibitem[Yi et~al\mbox{.}(2015)]%
        {yi2015understanding}
\bibfield{author}{\bibinfo{person}{Shuai Yi}, \bibinfo{person}{Hongsheng Li}, {and} \bibinfo{person}{Xiaogang Wang}.} \bibinfo{year}{2015}\natexlab{}.
\newblock \showarticletitle{Understanding pedestrian behaviors from stationary crowd groups}. In \bibinfo{booktitle}{\emph{Proceedings of the IEEE conference on computer vision and pattern recognition}}. \bibinfo{pages}{3488--3496}.
\newblock


\bibitem[Zheng and Liu(2019)]%
        {zheng2019improved}
\bibfield{author}{\bibinfo{person}{Shangfei Zheng} {and} \bibinfo{person}{Hong Liu}.} \bibinfo{year}{2019}\natexlab{}.
\newblock \showarticletitle{Improved multi-agent deep deterministic policy gradient for path planning-based crowd simulation}.
\newblock \bibinfo{journal}{\emph{Ieee Access}}  \bibinfo{volume}{7} (\bibinfo{year}{2019}), \bibinfo{pages}{147755--147770}.
\newblock


\bibitem[Zhou et~al\mbox{.}(2025)]%
        {zhou2025autovla}
\bibfield{author}{\bibinfo{person}{Zewei Zhou}, \bibinfo{person}{Tianhui Cai}, \bibinfo{person}{Seth~Z Zhao}, \bibinfo{person}{Yun Zhang}, \bibinfo{person}{Zhiyu Huang}, \bibinfo{person}{Bolei Zhou}, {and} \bibinfo{person}{Jiaqi Ma}.} \bibinfo{year}{2025}\natexlab{}.
\newblock \showarticletitle{AutoVLA: A Vision-Language-Action Model for End-to-End Autonomous Driving with Adaptive Reasoning and Reinforcement Fine-Tuning}.
\newblock \bibinfo{journal}{\emph{arXiv preprint arXiv:2506.13757}} (\bibinfo{year}{2025}).
\newblock


\bibitem[Zitkovich et~al\mbox{.}(2023)]%
        {zitkovich2023rt}
\bibfield{author}{\bibinfo{person}{Brianna Zitkovich}, \bibinfo{person}{Tianhe Yu}, \bibinfo{person}{Sichun Xu}, \bibinfo{person}{Peng Xu}, \bibinfo{person}{Ted Xiao}, \bibinfo{person}{Fei Xia}, \bibinfo{person}{Jialin Wu}, \bibinfo{person}{Paul Wohlhart}, \bibinfo{person}{Stefan Welker}, \bibinfo{person}{Ayzaan Wahid}, {et~al\mbox{.}}} \bibinfo{year}{2023}\natexlab{}.
\newblock \showarticletitle{Rt-2: Vision-language-action models transfer web knowledge to robotic control}. In \bibinfo{booktitle}{\emph{Conference on Robot Learning}}. PMLR, \bibinfo{pages}{2165--2183}.
\newblock


\end{thebibliography}





\end{document}